\documentstyle[12pt,epsf]{article}
\newcommand{\beq}{\begin{equation}}
\newcommand{\eeq}{\end{equation}}
\newcommand{\beqa}{\begin{eqnarray}}
\newcommand{\eeqa}{\end{eqnarray}}

\def\spose#1{\hbox to 0pt{#1\hss}}
\def\ltapprox{\mathrel{\spose{\lower 3pt\hbox{$\mathchar"218$}}
 \raise 2.0pt\hbox{$\mathchar"13C$}}}
\def\gtapprox{\mathrel{\spose{\lower 3pt\hbox{$\mathchar"218$}}
 \raise 2.0pt\hbox{$\mathchar"13E$}}}
\def\inapprox{\mathrel{\spose{\lower 3pt\hbox{$\mathchar"218$}}
 \raise 2.0pt\hbox{$\mathchar"232$}}}

\newcommand{\AmS}{{\protect\the\textfont2
  A\kern-.1667em\lower.5ex\hbox{M}\kern-.125emS}}

\begin{document}

\begin{titlepage}
\mbox{}\hfill FSU-SCRI-95-74\\
\begin{centering}
\vfill

{\bf SU(3) LATTICE GAUGE THEORY IN THE FUNDAMENTAL--ADJOINT PLANE AND
SCALING ALONG THE WILSON AXIS\footnote{Submitted to {\em Physics
Letters B}.}}

\vspace{1cm}
Urs~M.~Heller

\vspace{1cm}
{\em SCRI, The Florida State University,\\
Tallahassee, FL 32306-4052, USA\\}

\vspace{1.5cm}
{\bf Abstract}

\end{centering}

\vspace{0.3cm}\noindent
We present further evidence for the bulk nature of the phase transition
line in the fundamental--adjoint action plane of SU(3) lattice gauge
theory. Computing the string tension and some glueball masses along the
thermal phase transition line of finite temperature systems with $N_t=4$,
which was found to join onto the bulk transition line at its endpoint, we
find that the ratio $\sqrt{\sigma} / T_c$ remains approximately constant.
However, the mass of the $0^{++}$ glueball decreases as the endpoint of
the bulk transition line is approached, while the other glueball masses
appear unchanged.
This is consistent with the notion that the bulk
transition line ends in a critical endpoint, with the continuum limit there
being a $\phi^4$ theory with a diverging correlation length only in the
$0^{++}$ channel. We comment on the implications for the scaling behavior
along the fundamental or Wilson axis.

\vfill
\end{titlepage}

\section{Introduction}
\label{introduction}

The study of the phase diagram of fundamental--adjoint pure gauge systems,
\beq
S = \beta_f \sum_{P} [1 - \frac{1}{N} {\rm Re} {\rm Tr} U_{P} ]
+ \beta_a \sum_{P} [1 - \frac{1}{N^2} {\rm Tr} U^\dagger_{P} {\rm Tr}
U_{P}] ,
\label{eq:aaction}
\eeq
in the early 80's revealed a non-trivial phase structure with first order
(bulk) transitions in the region of small $\beta_f$
\cite{Greensite81,Bhanot82}. In particular a line of first order
transitions points towards the fundamental axis and, after terminating,
extends as a roughly straight line of bulk crossovers to the fundamental
axis and beyond. This non-trivial phase structure, and in particular the
critical endpoint, has been argued to be associated with the dip in the
discrete $\beta$-function of the theory with standard Wilson action, which
occurs in the region where the bulk crossover line crosses the fundamental
axis.

In a couple of recent papers Gavai, Grady and Mathur \cite{Gavai94} and
Mathur and Gavai \cite{Gavai94b} returned to investigating the behavior of
pure gauge SU(2) theory in the fundamental--adjoint plane at finite,
non-zero temperature. They raised doubts about the bulk nature of the phase
transition and claimed that their results were consistent with the
transitions, for lattices with temporal extent $N_t = 4$, 6 and 8, to be of
thermal, deconfining nature, displaced toward weak coupling with increasing
$N_t$. On the transition line for a fixed $N_t$ there is then a switch from
second order behavior near the fundamental axis to first order behavior at
larger adjoint coupling. In a Landau Ginzburg model of the effective action
in terms of Polyakov lines, Mathur \cite{Mathur95} reported that he could
reproduce the claimed behavior seen in the numerical simulations. These
results, should they be confirmed, are rather unsettling, since they
contradict the usual universality picture of lattice gauge theory with a
second order deconfinement transition for gauge group SU(2).

\begin{figure}
\begin{center}
\vskip 30mm
\leavevmode
\epsfysize=360pt
\epsfbox[90 40 580 490]{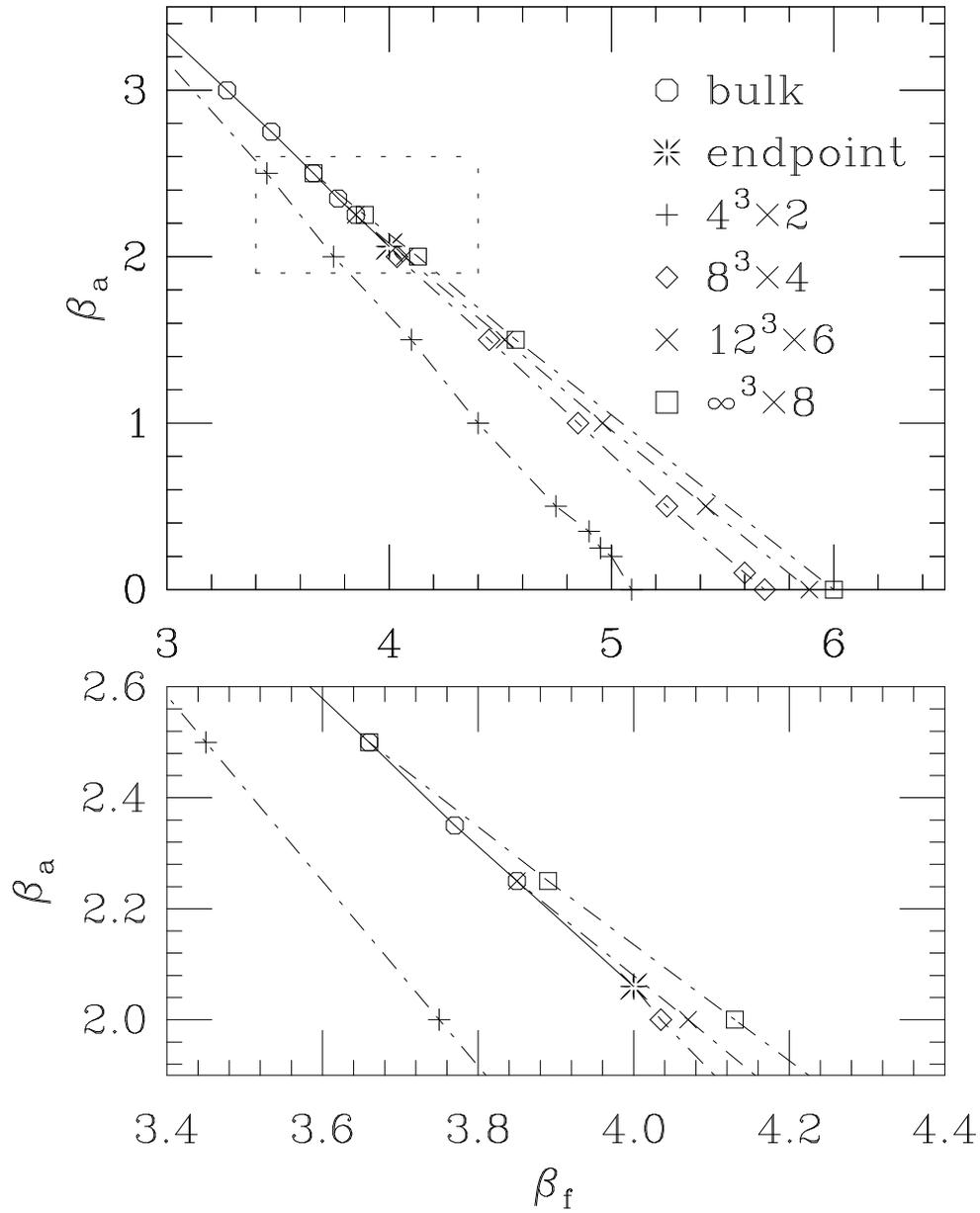}
\end{center}
\caption{The phase diagram together with the thermal deconfinement
         transition points for $N_t=2$, 4, 6 and 8 from
         Ref.~\protect\cite{SU3_F_Ab}. The lower plot shows
         an enlargement of the region around the end point of the bulk
         transition.}
\label{fig:phas_diag_T}
\end{figure}

Puzzled by the finding of Ref.~\cite{Gavai94}, we studied the finite
temperature behavior of pure gauge SU(3) theory in the fundamental--adjoint
plane \cite{SU3_F_A,SU3_F_Ab}. We obtained results in agreement with the
usual universality picture: there is a first order bulk transition line
ending at
\beq
     (\beta^*_f,\beta^*_a) = (4.00(7), 2.06(8)) .
\label{eq:endpoint}
\eeq
The thermal deconfinement transition lines for fixed $N_t$ (being of first
order for gauge group SU(3)) in the fundamental adjoint plane are
ordered such that the thermal transition for smaller $N_t$ occurs to the
left, at smaller $\beta_f$, than that for a larger $N_t$. In this order the
thermal transition lines join on to the bulk transition line. The thermal
transition line for $N_t = 4$ joins the bulk transition line very close to
the critical endpoint. This is shown in Figure 4 of Ref.~\cite{SU3_F_Ab},
reproduced here as Figure~\ref{fig:phas_diag_T}.

To solidify this picture, which is in agreement with the usual universality
scenario, we have continued the investigation studying zero-temperature
observables. We computed the string tension and the masses of some
glueballs, in particular the $0^{++}$ glueball, along the thermal transition
line for $N_t = 4$. For universal continuum behavior $\sqrt{\sigma}$ and
the glueball masses should be constant along the thermal transition line for
a fixed $N_t$, leading to constant ratios $T_c/\sqrt{\sigma}$ and
$m_g/\sqrt{\sigma}$.

Of course, at small $N_t$, corresponding to a large lattice spacing $a$,
we expect to see some deviations from this constant behavior. However, we
find large deviations for $m_{0^{++}}/\sqrt{\sigma}$ as we approach the
critical endpoint of the bulk transition along the $N_t=4$ thermal
transition line. The scalar glueball mass decreases dramatically, much more
than what could be expected from simple scaling violations at a large lattice
spacing. On the other hand, this is not really a surprise since at the
critical endpoint at least one mass in the $0^{++}$-channel has to vanish.
We elaborate on our findings in the next sections and then discuss the
implications for the scaling behavior along the fundamental (or Wilson) axis.

\section{Observables and Analysis}
\label{analysis}

We have made simulations of the model with action (\ref{eq:aaction}) along
the thermal transition line for $N_t=4$, and continued along the bulk
transition line, on a $12^4$ lattice. Experience
indicates this size to be sufficient to avoid significant finite size
effects. The simulations were carried out with a 10-hit Metropolis
algorithm tuned to an acceptance rate of about 50\%. Observables were
measured every 20 sweeps after thermalization. Close to the fundamental
axis this resulted in essentially statistically independent measurements,
while closer to the critical endpoint the autocorrelation time was
significantly increased. We computed finite $T$ approximants to the potential
from time-like Wilson loops constructed with `APE'-smeared spatial links
\cite{APE_smear} to increase the overlap to the ground state potential. On
and off axis spatial paths were considered with distances $R = n$,
$\sqrt{2} n$, $\sqrt{3} n$ and $\sqrt{5} n$, with $n = 1$, 2, \dots an
integer. The string tension was then extracted from the usual fit
\beq
V(\vec R) = V_0 - \frac{e}{R} + l \left( G_L(\vec R) - \frac{1}{R} \right)
 + \sigma R .
\label{eq:fit_form}
\eeq
Here $G_L(\vec R)$ is the lattice Coulomb potential, included in the fit
to take account of short distance lattice artefacts. Our fits are fully
correlated $\chi^2$-fits with the correlations estimated by bootstrap,
after binning to alleviate autocorrelation effects. The results
of the best fits are listed in Table~\ref{tab:fits}.

\begin{table}[ht]
\centerline{%
\begin{tabular}{|l|l|l|l|r|l|l|c|} \hline
 $\beta_a$ & $\beta_{fc}(N_t=4)$ & $\beta_f$ & $L$ & $N_{meas}$ &
 $\sqrt{\sigma}$ & $m_{0^{++}}$ & $m_{2^{++}}(t=1)$ \\ \hline
 0.0  & 5.6925(2)  & 5.7$^a$    &    &      & 0.4099( 12) & 0.97( 2) &
 2.39(13) \\
 0.5  & 5.25(5)    & 5.25       & 12 &  500 & 0.4218( 28) & 0.93(11) &
 2.29(13) \\
 1.0  & 4.85(5)    & 4.85       & 12 &  500 & 0.4024( 82) & 0.78(28) &
 2.45(19) \\
 1.5  & 4.45(5)    & 4.45       & 12 &  500 & 0.3743( 51) & 0.56(17) &
 2.13( 9) \\
 2.0  & 4.035(5)   & 4.03       & 12 & 1000 & 0.555 ( 11) & 0.34( 6) &
 3.11(16) \\
 2.0  & 4.035(5)   & 4.035      & 12 & 2000 & 0.4725(128) & 0.20( 4) &
 3.17(14) \\
 2.0  & 4.035(5)   & 4.04       & 12 & 1000 & 0.3750( 24) & 0.27( 8) &
 2.37( 9) \\ \hline
 2.25 & 3.8475(25) & 3.8475$^b$ & 12 &  500 & 0.619 ( 18) & 0.37(10) &
 3.15(37) \\
 2.25 & 3.8475(25) & 3.8475$^c$ & 12 &  500 & 0.3005( 22) & 0.61( 8) &
 1.59( 7) \\
 2.25 & 3.8475(25) & 3.8475$^c$ & 16 &  500 & 0.2965( 19) & 0.62( 5) &
 1.72( 9) \\ \hline
\end{tabular}%
}
\caption{The results in the neighborhood of the $N_t=4$ thermal transition
         line. Comments: (a) $\protect\sqrt{\sigma}$ at $\beta_a=0.0$
         comes from \protect\cite{MTc_sig}, $m_{0^{++}}$ from
         \protect\cite{GF11_gb}; we did not take their best value, but rather
         the effective mass from the same distance as at $\beta_a=0.5$;
         $m_{2^{++}}$ is from \protect\cite{dFSST86}; (b) in the disordered
         phase and (c) in the ordered phase on the bulk transition.}
\label{tab:fits}
\medskip\noindent
\end{table}

We also computed glueball correlation functions in the $0^{++}$, $2^{++}$
and $1^{+-}$ channel that can be built from simple plaquette operators. In an
attempt to improve the signals we built these plaquettes not only from the
original, but also from the smeared links, already used for the computation
of the potential. Not surprisingly for computations done around the
critical coupling for the $N_t = 4$ thermal phase transition, we did not
obtain a significant signal in the $1^{+-}$ channel and only an effective
mass from time distances $t = 0/1$ in the $2^{++}$ channel -- we had 500
measurements everywhere, except for $\beta_a = 2.0$, near the critical
endpoint where the number was increased, as given in Table~\ref{tab:fits}.
In the $0^{++}$ channel we got a signal at distance $t = 1/2$ at small
$\beta_a$ and out to $t = 3/4$ at $\beta_a = 2.0$. Our best results are also
given in Table~\ref{tab:fits}.

\begin{figure}
\begin{center}
\leavevmode
\epsfysize=360pt
\epsfbox[90 40 580 490]{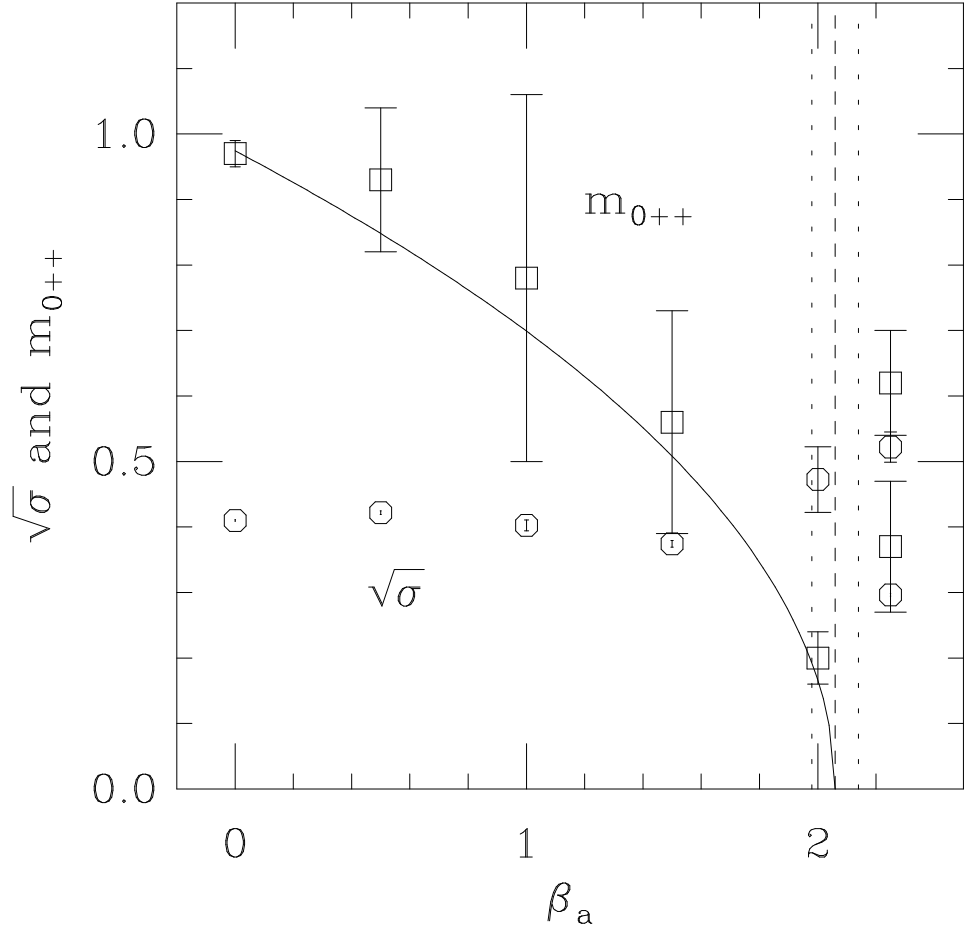}
\end{center}
\caption{$\protect\sqrt{\sigma}$ (octagons) and $m_{0^{++}}$ (squares) as
         a function of $\beta_a$ along the thermal transition line for
         $N_t=4$. At $\beta_a=2.25$ we show the results from both phases
         at the bulk transition. The dashed vertical line gives the
         location of the critical endpoint, (\protect\ref{eq:endpoint}),
         with the dotted lines indicating the error band. The curve is a
         fit to $m_{0^{++}}=A(2.06-\beta_a)^{1/2}$.}
\label{fig:m_and_sig}
\end{figure}

The quantities $\sqrt{\sigma}$ and $m_{0^{++}}$ are shown in
Figure~\ref{fig:m_and_sig} plotted versus $\beta_a$. As can be seen,
$\sqrt{\sigma}$ remains approximately constant along the thermal transition
line for $N_t=4$ -- the errors shown in the figure are statistical only; no
estimate of the error from the uncertainty in the determination of
$\beta_{fc}$ has been attempted except for $\beta_a=2.0$. There, the
computation has been repeated for two nearby couplings, also listed in
Table~\ref{tab:fits}; the variation with $\beta_f$ becomes so rapid that
the error in the determination of $\beta_{fc}$ becomes the dominating factor.

While our estimate for $m_{2^{++}}$, albeit an unreliable estimate since we
had to use distances $t=0$ and 1 to obtain enough of a signal to extract an
effective mass, also remains approximately constant, $m_{0^{++}}$ shows a
remarkable decrease as the critical endpoint is approached. This observed
behavior suggests that the mass in the $0^{++}$ channel vanishes at the
critical endpoint, thereby giving strong additional evidence for the
existence of this critical endpoint, since at a critical point at least one
mass, in the $0^{++}$ channel to have a rotationally invariant continuum
limit, must vanish.

Since no other observable seems to be dramatically affected by the critical
endpoint, we conjecture that the continuum theory one would obtain there is
simply the (trivial) $\phi^4$ theory. To substantiate this claim somewhat,
we made a fit to the scalar mass of the form
\beq
m_{0^{++}} = A \left( \beta^*_a - \beta_a \right)^p
\label{eq:massfit}
\eeq
expected near a critical point. A 3-parameter fit gave $A=0.76(11)$,
$p=0.35(20)$ and $\beta^*_a=2.02(6)$ with a $\chi^2$ of $0.29$ for 2 dof.
Note that the estimate for $\beta^*_a$ is in agreement with the previous
estimate (\ref{eq:endpoint}) obtained in \cite{SU3_F_Ab} from fits to the
jump in the plaquette across the bulk transition line. Within its large
error, the exponent $p$ is compatible with the mean field value $0.5$ of
$\phi^4$ theory, up to logarithmic corrections. Since the errors of the
fit parameters are rather large we also made a fit with $\beta^*_a$ held
fixed at its value $2.06$ of (\ref{eq:endpoint}). This fit gave $A=0.71(3)$
and $p=0.44(5)$ with a $\chi^2$ of $0.47$ for 3 dof. Again, $p$ is
compatible with the mean field value. Finally, a fit with $\beta^*_a=2.06$
and $p=0.5$ both held fixed gave $A=0.68(1)$ with a $\chi^2$ of $1.50$ for
4 dof. This last, still very acceptable fit is included in
Figure~\ref{fig:m_and_sig}.

\begin{figure}
\begin{center}
\leavevmode
\epsfysize=360pt
\epsfbox[90 40 580 490]{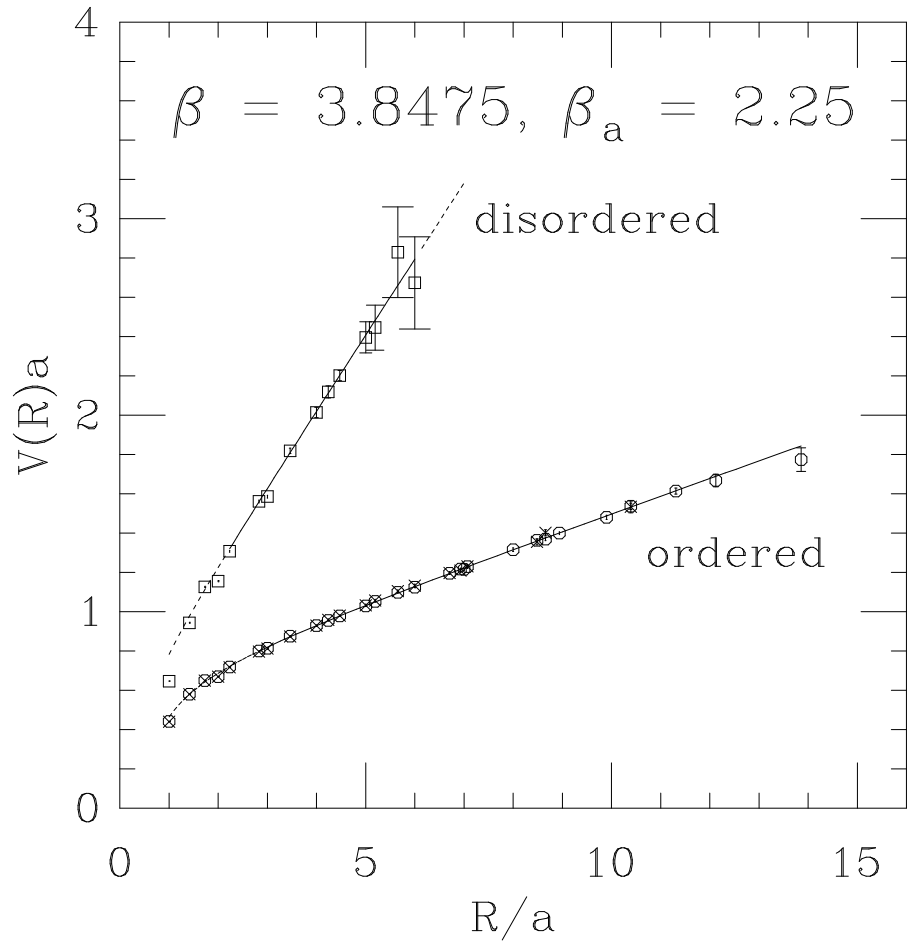}
\end{center}
\caption{The potential in both phases on the bulk transition line at
         $(\beta_f,\beta_a)=(3.8475,2.25)$ on $12^4$ lattices ($\Box$ and
         $\times$) and on a $16^4$ lattice ($\circ$).}
\label{fig:pot_ba2p25}
\end{figure}

While we believe to have assembled impressive further evidence for the bulk
nature of the phase transition line in the fundamental--adjoint plane and of
its critical endpoint, we decided to make one further test. We computed the
potential in both phases on the bulk transition line at $\beta_a = 2.25$.
This is about the location where the $N_t=6$ thermal transition line joins
onto the bulk transition line according to Ref.~\cite{SU3_F_Ab}. Hence, even
in the ordered phase a lattice of size $12^4$ should be large enough to
obtain a reliable potential. To make sure that this is indeed the case, we
repeated the computation in the ordered phase on a $16^4$ lattice. As can
be seen in Table~\ref{tab:fits} the finite size effect on the $12^4$
lattice does indeed appear negligible. On both sides of the bulk transition
-- at identical couplings: we did not observe even an attempt at tunneling
from one phase to the other -- we find a clearly confining potential.
Therefore the bulk transition does not, as expected, affect confinement,
provided the lattices are large enough.

Of course, the string tension (and glueball masses) jump from one side of
the bulk transition to the other. Indeed, as the bulk transition line is
approached from the fundamental axis the string tension varies more and
more rapidly -- and the thermal deconfinement transition lines for different
$N_t$ come closer together -- {\it i.e.,} the dip in the step $\beta$-function
becomes deeper until a jump is developed through the bulk transition line.

\section{Implications for Scaling}
\label{scaling}

In the previous section we have corroborated the existence of a first order
bulk phase transition line ending in a critical endpoint. We have provided
evidence that physical observables are little affected by this phase
transition line, except for a stronger dependence on $\beta_f$ at fixed
$\beta_a$ -- a deepening of the dip in the step $\beta$-function -- and
eventually a jump across the bulk transition line. The notable exception to
this is the glueball mass in the $0^{++}$ channel which decreases as the
critical endpoint is approached and vanishes there.

\begin{figure}
\begin{center}
\leavevmode
\epsfysize=360pt
\epsfbox[90 40 580 490]{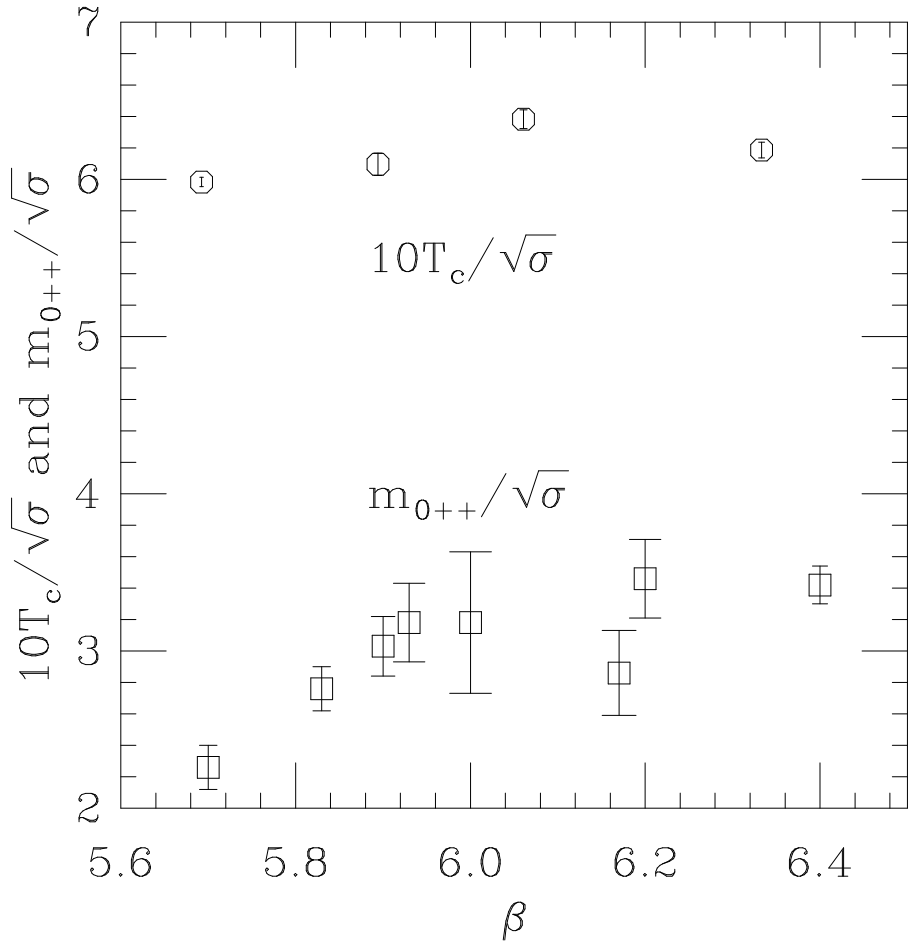}
\end{center}
\caption{$10T_c/\protect\sqrt{\sigma}$ (octagons) and
         $m_{0^{++}}/\protect\sqrt{\sigma}$ (squares) as a function of
         $\beta$ for the fundamental (Wilson) action.}
\label{fig:m0_ov_sig}
\end{figure}

The influence of the critical endpoint on the $0^{++}$ glueball appears
still visible in the crossover region on the fundamental (Wilson) axis.
This has first been argued in Ref.~\cite{Rajan86}. The $0^{++}$ glueball is
lighter in the crossover region, leading to a different scaling behavior
than other observables. This can be seen in Figure~\ref{fig:m0_ov_sig}
where we show the latest data of $T_c/\sqrt{\sigma}$ from Ref.~\cite{Boyd95}
and $m_{0^{++}}/\sqrt{\sigma}$ with $\sqrt{\sigma}$ taken from
Refs.~\cite{MTc_sig,B-S_sig} and the glueball mass from
Refs.~\cite{GF11_gb,M-T_gb,Bali_gb}. While $T_c/\sqrt{\sigma}$ stays
approximately constant in the $\beta$ interval shown,
$m_{0^{++}}/\sqrt{\sigma}$ decreases visibly in the crossover region around
$\beta = 5.7$.

It has long been known that the scalar glueball mass scales differently in
the crossover region than $T_c$ and the string tension, whose scaling
behavior as a function of $\beta$ deviates from asymptotic scaling but
agrees with the step $\beta$-function found in MCRG computations. The
scalar glueball seemed more compatible with asymptotic scaling. However, in
view of our findings that the different behavior of the scalar glueball
mass comes from the influence of the nearby critical endpoint this
asymptotic scaling behavior appears to be accidental. It would be
interesting to determine the scaling behavior of other (pure gauge)
observables. We conjecture that they would behave more like $T_c$ or
$\sqrt{\sigma}$ than like the scalar glueball mass. Unfortunately no
reliable, large volume, glueball masses for $J^{PC}$ quantum numbers other
than $0^{++}$ for couplings in the crossover region exist in the literature
to check this conjecture.

\section*{Acknowledgements}

This work was partly supported by the DOE under grants
\#~DE-FG05-85ER250000 and \#~DE-FG05-92ER40742. The computations were
carried out on the cluster of IBM RS6000's at SCRI.


\end{document}